# Electron transport in bilayer graphene nano constrictions patterned using AFM nanolithography

Robert W. Rienstra[1], Nishat Sultana[1], En-Min Shih[4], Evan Stocker[1], Kenji Watanabe[2], Takashi Taniguchi[3], Curt A. Richter[4], Joseph Stroscio[4], Nikolai Zhitenev[4] and Fereshte Ghahari[1][†]

[1]Department of Physics and Astronomy, George Mason University; Fairfax, VA 22030, USA.
[2]Research Center for Electronic and Optical Materials, National Institute for Materials Science; 1-1 Namiki, Tsukuba, 305-0044, Japan.
[3]Research Center for Materials Nanoarchitectonics, National Institute for Materials Science; 1-1 Namiki, Tsukuba, 305-0044, Japan.
[4]Physical Measurement Laboratory, National Institute of Standards and Technology; Gaithersburg, MD 20899, USA.

[†]Corresponding author: fghahari@gmu.edu

**ABSTRACT:**

Here we report on low temperature transport measurements of encapsulated bilayer graphene nano constrictions fabricated employing *electrode-free* AFM-based local anodic oxidation (LAO) nanolithography. This technique allows for the creation of constrictions as narrow as 20 nm much smaller than previous studies. In wider constrictions, we observe bulk transport characteristics. However, as the constriction's width is reduced, a transport gap appears. Single quantum dot (QD) formation is observed within the narrowest constriction with addition energies exceeding 100 meV, which surpass previous experiments on patterned QDs. Our results suggest that transport through these narrow constrictions is governed by edge disorder combined with quantum confinement effects. Our findings introduce *electrode-free* AFM-LAO lithography as an easy and flexible method for creating nanostructures with tunable electronic properties without relying on patterning techniques such as e-beam lithography. The excellent control and reproducibility provided by this technique opens exciting opportunities for carbon-based quantum electronics and spintronics.

**Main text:**

Understanding quantum electronic properties of confined charge carriers in graphene-based nano structures has recently attracted increasing attention mainly due to their potential applications for the development of solid-state spin qubits and innovative future devices [1]. For example, the low nuclear spin concentration and the weak spin-orbit interaction in graphene-based quantum dots (QDs) lead to longer spin coherence times compared to semiconductor qubits.

Recent advancements have enabled the fabrication of quantum point contacts (QPCs) and quantum dots (QDs) in both graphene and bilayer graphene employing various techniques, including split gates, e-beam lithography, etching and cutting [2-26]. In bilayer graphene, these confined geometries have been mostly created by charge depletion using split gates where an application of a transverse electric field opens a band gap and thus depletes electrons in gated areas [10-14, 48, 49]. In principle, a similar split gate geometry cannot induce a complete charge depletion in graphene due to Klein tunneling of Dirac electrons which results in high transparency of barriers at zero incident angle [15-17]. A complete confinement at zero magnetic field can be only obtained by removal of graphene, e.g. by etching [18-24]. However, many of these commonly used techniques are time-consuming as they require multiple steps of stacking, lithographic patterning and reactive ion etching (RIE). Furthermore, etching byproducts, residuals, and edge roughness can create charge localization and adversely affect electronic performance. In contrast, electrostatic gate defined geometries have been shown to preserve device quality at the cost of increased complexity, imperfect confinement, and imposing the requirement of a tunable bandgap. An alternate technique for nanopatterning is cutting the graphene sheet. For example, in a recent experiment, an AFM-based technique has been employed to cut graphene into very narrow constrictions where conductance quantization was observed [25]. However, for this method to work, the mechanical cleavage needs to be along high symmetry crystallographic directions which can limit controllability of creating such kind of constrictions. Another technique is AFM-based local anodic oxidation (LAO). Although this method has been employed before, its application is seriously limited due to the required microelectrode connections for application of a DC voltage between the tip and the sample [26-30, 44].

Here we report on fabrication and transport measurements of mechanically defined bilayer graphene nano constrictions utilizing the recently developed *electrode-free* AFM-based LAO nanolithography [31]. Using this technique, we have fabricated constrictions as narrow as 20 nm and studied a range of widths up to 125 nm employing low temperature transport measurements. While larger constrictions exhibit an energy gap, we observe quantum dot (QD) transport characteristics in 30 nm wide constrictions at low temperatures. We employ bias and magnetic field spectroscopy measurements to characterize these nano constrictions. This introduces *electrode-free* AFM-LAO lithography as a unique technique for on demand fabrication of graphene-based nano constrictions.

In *electrode-free* AFM-LAO lithography, graphene exfoliated onto a Si/SiO$_2$ wafer is placed in a high humidity environment [31]. A relatively high AC voltage is applied to the AFM tip which couples to the substrate to form and stabilize a water meniscus, a small water droplet, between the tip and the sample. The water meniscus only forms when the relative humidity around the AFM is sufficiently high. When the AC voltage becomes greater than a threshold voltage, the water dissociates into H+ and OH- groups. The OH- oxidizes the graphene and then a force as low as ~10 nN is applied making a cut in the graphene sheet (Fig. 1a). By moving the tip under these conditions, complex patterns can be etched into the graphene layer. The quality and width of an individual cut can vary widely depending on the applied voltage, writing speed, and the individual AFM probe among other factors. Once the lithography parameters were optimized, nano constrictions were cut into multiple mono and bilayer graphene sheets (Figs. 1b, c). These constrictions are designed such that their widths are much smaller than their length to avoid creating long nanoribbons. The final flakes were then vacuum annealed to eliminate the water vapor, and any other contaminants adsorbed prior to device incorporation. Using a dry pickup technique, the annealed nano- constrictions were encapsulated with hexagonal boron nitride (hBN) to preserve the electronic quality of the graphene through subsequent nanofabrication processes. As with vacuum annealing, the pickup process, and subsequent fabrication steps, can result in the destruction of several nano-constrictions. Due to these many challenges inherent in assembling pre-patterned graphene nano-constrictions into van der Waals (vdW) heterostructures, we also applied *electrode-free* AFM-LAO nanolithography to pre-assembled hBN encapsulated samples. During this process the AC excitation voltage actively couples to the graphene underneath. This provides a sufficiently high electric field to dissociate the top layer of hBN and remove both the top layer hBN and graphene below. Fig. 1c displays an AFM image of a pre-assembled graphene heterostructure for which two narrow constrictions were etched after encapsulation. By performing *electrode-free* AFM-LAO after encapsulation the fabrication process is greatly simplified. The risk from encapsulation, vacuum annealing, and contamination are eliminated resulting in a higher device yield. We have studied constrictions of both types in this work (Fig. 1).

After cutting and encapsulation, the heterostructures are shaped into a Hall bar geometry employing reactive ion etching, where the constriction connects to two large graphene reservoirs which can be considered as external contacts being coupled to the central narrow constriction. Finally, the electrical edge contacts were made to the conducting channels (Figs. 1e, f).

Transport measurements have been performed employing standard lock-in techniques at a frequency of 17.77 Hz and with a small AC excitation voltage ($\leq 100\,\mu$V). In this method, the conductance of the constriction is measured by biasing the graphene reservoir electrodes. Figures 2a, b, c show the conductance $G$ as a function of back gate voltage, $V_g$, for three representative bilayer graphene constrictions with different widths. The constriction with width of $W=125 \pm 5$ nm (Device P1) exhibits the behavior typical of bulk graphene as shown in Fig. 1a. Additionally several kinks with quantization close to integer multiples of $e^2/h$ are observable on the electron side suggestive of quantum size effects expected in a QPC. The constriction with a smaller $W=100 \pm 5$ nm (device P2), however, exhibits a relatively small gap at the charge neutrality point where the conductance is reaching zero near the charge neutrality point (Fig. 2b). This suppression is due to the so-called transport gap which has been previously reported for nanoribbons [32-35, 44]. The constriction with an intermediate width of $75 \pm 5$ nm (device P3) shows a bigger transport gap with conductivity being suppressed in a large gate range (-2 V< $V_g$ <15 V) as shown in Fig. 2c. This agrees with previous studies on graphene nanoribbons where the gap scales inversely with the width of constriction [32].

We can obtain a direct estimate of the size of the observed energy gap by measuring the differential conductance as a function of both the back gate and bias voltage. Figure 2c displays the differential conductance ($G=dI/dV_b$) measured at $T = 1.6$ K as a function of the bias voltage, $V_b$, and back gate voltage $V_g$ for device P3, which shows a relatively large transport gap. A gapped region specified by a dark blue area is observed in the $V_g$ -$V_b$ plane where conduction and valence bands lie outside the bias window. The band gap, $E_g = eV_b$, can be evaluated from the value of the bias voltage at the vertex of the gapped region giving an estimate of $E_g \approx 50$ meV. In principle, in these narrow constrictions, the geometric quantum confinement can result in an energy gap. Based on a simple quantum mechanical model the gap for a bilayer graphene constriction can be expressed as $E_q \approx \pi^2 \hbar^2 / m^* w^2$, where $w$ is the channel width and $m^* \sim 0.033 m_c$ is the effective mass in bilayer graphene with $m_c$ the charged carrier mass. This gives an estimated gap of $E_q \approx 2.8$ meV. Moreover, the vertical displacement field, $D$, across the two carbon layers can give rise to an energy gap. $D$ can be calculated from the value of back gate voltage $V_g$ at the center of the gapped region by the formula, $D = (-C_g V_g)/2\varepsilon_0$, where $C_g$ is the capacitance of bottom hBN layer, giving an estimate of $\approx 33$ mV/nm and the corresponding energy gap of $E_D \approx 3.3$ meV. These calculations give a total energy gap of $\approx 6$ meV which is much smaller than the experimentally observed energy gap. This suggests that the enhanced energy gap is induced by other mechanisms such as impurity and edge disorder effects as previously speculated for graphene and bilayer graphene nanoribbons [34, 45-47, 50-51]. As our constrictions are encapsulated with hBN, we rule out the charge impurity and electron-hole puddles for this large enhancement of energy gap. This enhancement is likely caused by edge disorder effects [47]. AFM etching can be a more violent process that leaves behind rough edges. Furthermore, the biproducts of hBN dissociation when exposed to the high humidity environment can introduce additional disorder at the edges [36, 37]. Based on theoretical investigations, this roughness and disorder can create charge localization along the edge and nanoscale size regions where conductance is governed by transport through multiple dots [47].

Now, we discuss the transport characteristics of the narrowest constriction. Figure 3a shows conductance versus back gate voltage $V_g$ for Device P4 with $W=30 \pm 5$ nm. We observe a series of clear periodic conductance maxima as a function of $V_g$ characteristic of Coulomb blockade peaks, suggesting single QD formation within this narrow constriction as discussed below. In traditional QDs patterned by ebeam lithography, confinement is typically achieved by connecting a graphene island to source and drain electrodes through two narrow constrictions [3, 33]. However, charge localization can also be achieved without creating these additional constrictions. This type of QD formation has been previously observed in graphene nanoribbons on $SiO_2$ and was attributed to edge disorder and charge inhomogeneities in $SiO_2$ combined with quantum confinement effects [33, 34, 38]. As for the observed enhanced energy gap in wider constrictions, we believe the QD formation in this narrowest hBN encapsulated constriction is a result of the charge localization by edge disorder and possibly potential fluctuations caused by external dangling bonds along the edge [39, 47]. When the edge roughness is comparable to the width, single or multi QD formation is expected depending on the channel's dimensions [47].

For a detailed analysis of the QD characteristics observed in device P4, we carried out bias spectroscopy measurements at $T$= 1.6 K. Figure 3b shows the differential conductance ($G=dI/dV_b$) measurements as a function of the bias voltage, $V_b$, and back gate voltage $V_g$. The corresponding periodic Coulomb diamonds are visible with the strong modulation of the energy gaps, the so-called addition energies, ranging from 10 meV to well above 100 meV. The maximum addition energies are much larger than those in the previous experiments on patterned structures indicating that the size of the QD is much smaller as will be discussed below. In Fig. 3b, the Coulomb diamonds are more pronounced at higher negative gate voltages while at lower gate voltages the regions of multi dot behavior are observable likely due to lesser screening and potential fluctuations. The same device on the second cool down exhibit more pronounced Coulomb diamonds in all density ranges as shown in Fig. S1.

The bias spectroscopy measurements allow us to extract the single particle energy levels of the QD. The separation in gate voltage between the successive Coulomb resonances as shown in Figs. 3a, b is related to the energy required to add the next charged particle to the dot. This energy is approximately equal to the sum of the charging energy (due to coulomb interaction) and the separation between the successive single particle energy levels. For two degenerate single particle energy levels of the QD, adding an electron only costs the charging energy, which is assumed to be independent of the magnetic field. Thus, the minimal separation between Coulomb peaks in back gate voltage corresponds to the charging energy. To find the charging energies, we use the differential conductance bias measurements to determine the back gate lever arm for each electron being added to the dot [40, 41]. Using these lever arms, the gate voltage axis can be converted into energy such that the $\Delta E = \alpha e \Delta V_g$, where $\Delta E$ and $\Delta V_g$ are the gate voltage and energy separation between coulomb resonances and $\alpha$ is the lever arm. The addition energies for adding electrons to the quantum dot are extracted from Figs. 3a, b and are plotted in Figs. 3c. This energy generally decreases by adding more electrons to the dot as expected since the electronic size of the dot increases. In Fig. 3c, we observe clear enhancement of addition energy as the dot is filled with the four and eight electrons. This is expected as single particle energy levels have a four-fold spin and valley degeneracy in bilayer graphene [42]. The same calculations for second cool down also show a near four-fold degeneracy (Fig. 3d).

Using data in Fig. 3, we can now obtain an estimate of the size of QD. One way relies on comparing the gate capacitance of the dot, $C_g$, extracted from the average energy separation in gate between coulomb peaks given as $\Delta E_{ave} = \frac{e^2}{C_g}$ to the gate capacitance per unit area, $\bar{C}_g$, of the device as determined from the magnetic field measurements outside the constriction. In the quantum hall regime, the gate capacitance can be expressed as $\frac{e^2 B \nu}{h} = \bar{C}_g (V_g - V_{NP})$, where $\nu$ is the occupation number and $V_{NP}$ is the location of the charge neutrality point. Employing this relation for various filling factors, the gate capacitance per unit area was determined to be $\bar{C}_g = 119.73$ $aF/\mu m^2$. Then the total area of the dot can be estimated by $A = \frac{C_g}{\bar{C}_g}$, where its diameter is given by $a = \sqrt{\frac{4 \frac{C_g}{\bar{C}_g}}{\pi}}$. This relation gives an estimate of 17.48 nm and 26.38 nm for the dot's diameter for the first and second cool downs, respectively (see also supplementary section A).

Overall, these calculations give an estimate for the size of the QD which is close to actual geometric size of constriction obtained from AFM measurements. This suggests that transport in our narrowest constriction is through a single isolated island. The estimated size of the single QD is smaller for first cool down compared to second one. This is likely due to a different disorder configuration and less visibility of Coulomb diamonds for data presented in Fig. 3a. To improve these estimations, a theoretical model which includes the non-radial symmetry of the QD, and the nonhomogeneous confinement potential should be considered.

We now study the transition from electrostatic confinement to magnetic confinement in our quantum dots by looking at the shift of single particle energy levels as a function of magnetic field, $B$. Figure 4a (also see Fig. S2) shows the differential conductance measurements of Coulomb resonances as a function of $V_g$ and $B$. The conductance peak positions in $V_g$ and $B$ are extracted from Fig. 4a and plotted separately in Fig. 4b. By subtracting the charging energies between neighboring resonances which is not dependent on field, we can extract the single particle quantum energy levels of the quantum dot at each magnetic field. The dispersion of these energy levels with magnetic field is displayed in Fig. 4c. A four level bunching of quantum energies is observed as expected. Qualitatively, the common trend observed for single particle energies is that lower energy levels slightly bend toward higher energies, while the highest filled level does not shift appreciably with increasing *B*. This is consistent with previous measurements, where similar behavior was attributed to the interplay between the confinement and Landau level (LL) formation due to finite size of the island [43]. Moreover, we observe splitting of energy levels with field, where one pair shifting down, the other slightly moving up in energy. The observed dispersions are suggestive of valley splitting observed in previous experiments [14], however higher magnetic fields are required to confirm this further. Overall, our data shows the single particle energy levels do not disperse appreciably with field at least up to fields of the order of 3 to 4 T. In principle, in the presence of a magnetic field, electrons are experiencing the cyclotron orbit motion being confined to an area with radius $l_B = \sqrt{\frac{\hbar}{eB}}$. A transition from QD states to LLs is expected to occur when the magnetic field length scale $l_B$ is comparable to the confinement length scale which can be estimated as the radius of the QD. The magnetic length $l_B$ is expected to be around 8 nm to 14 nm by increasing the field from 3 T to 9 T. This give an estimate for the diameter of the island to be on the order of 16 nm to 28 nm, relatively close to its geometric size consistent with previous calculations.

In conclusion, we have shown that AFM-based LAO nanolithography can be employed to fabricate nano constrictions with various width in graphene-based vdW heterostructures. The electronic properties of these constrictions can be tuned by changing their r width. In wider constrictions, we observe bulk transport characteristics. However, a transport gap is observed as the constriction's width is reduced. The narrowest constriction exhibit QD behavior where Coulomb peaks are visible in all gate ranges with addition energies exceeding 0.1 eV which is larger than previous experiments on patterned structures. Our bias and field spectroscopy measurements suggest that transport in our narrowest constriction is through a single isolated island. Our results indicate that transport through these nano constrictions is governed by edge disorder combined with quantum confinement effects. This introduces AFM lithography as a unique technique for on demand production of graphene nano constrictions where the electronic properties can be tuned by changing the width. This is promising for their possible applications for designing on chip electronic circuits and the development of solid-state spin qubits.

**Author Contributions:** RR fabricated devices and acquired and analyzed experimental data supervised by FG. NS and ENS helped with device fabrication and AFM lithography, respectively. ES helped with data analysis. TT and KW grew the hBN crystals. All authors contributed to writing the paper.

**Acknowledgments:** We thank Akhil Chauhan for preliminarily device characterization. RR, NS and ES thank Department of energy (DOE), award # DE-SC0024109. RR and NS thank the CNST for their support for device fabrication. T.T. and K.W. acknowledge support through Japan Society for the Promotion of Science KAKENHI Grants No. 21H05233 and 23H02052 and World Premier International Research Center Initiative (WPI), Ministry of Education, culture, sports, science and technology (MEXT), Japan.

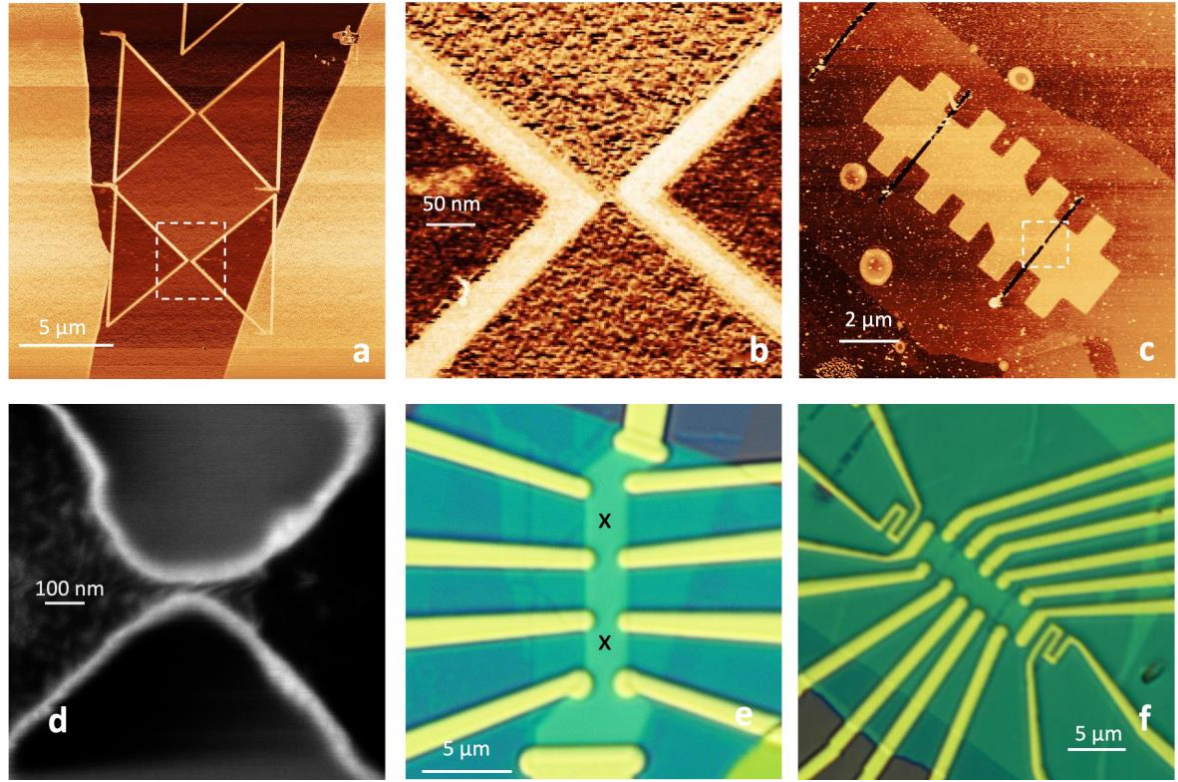

**FIG. 1: Fabrication of graphene-based nano constriction devices employing *electrode free* AFM-LAO nanolithography**. (**a**) AFM image of bilayer graphene etched into narrow constrictions by AFM nanolithography. (**b**) Zoomed in AFM image of the area inside the rectangle in (a) showing a constriction with a width of $W=30 \pm 5$ nm. (**c**) AFM image of a pre-assembled bilayer graphene etched into narrow constrictions by AFM nanolithography after hBN encapsulation. The area inside the rectangle shows the nano constriction. (**d**) Scanning electron microscopy (SEM) image of a graphene nano constriction with a width of $W=20 \pm 5$ nm. (**e**) Optical images of the final device with electrical contacts made from the constriction in (b). The "X" shows the location of the constrictions (**f**) Optical images of the final device with electrical contacts made from the constriction in (c).

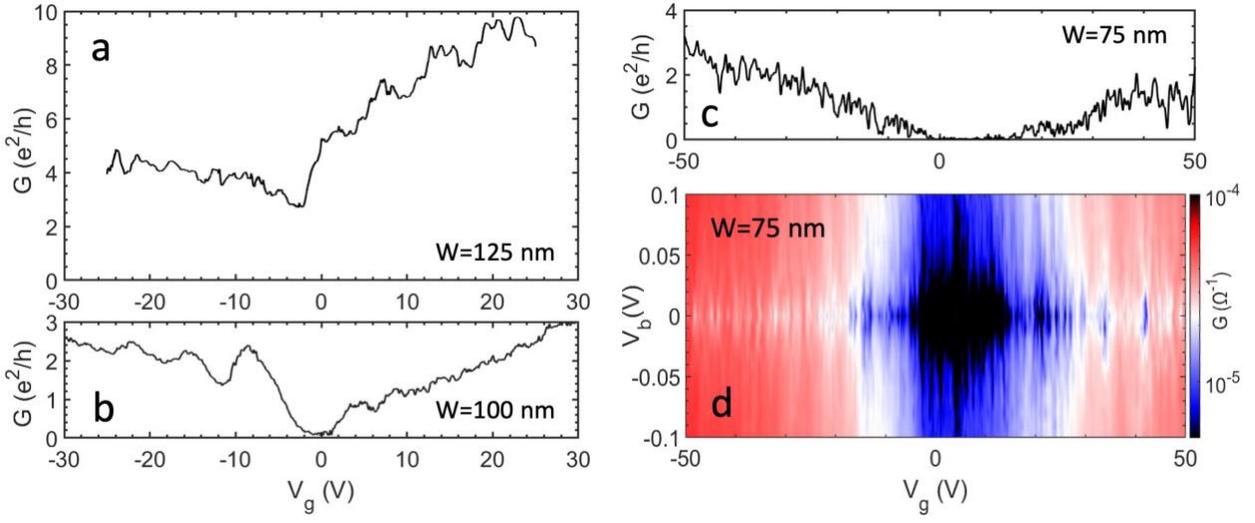

**FIG. 2: Transport measurements of bilayer graphene nano constrictions with varying width.** Conductance versus $V_g$ measured for (**a**) device P1 with $W=125 \pm 5$ nm (**b**) device P2 with $W=100 \pm 5$ nm, (**c**) device P3 with a width of $W=75 \pm 5$ nm measured at $T= 1.6$ K. (**d**) The differential conductance ($G=dI/dV_b$) as a function of the bias voltage, $V_b$, and back gate voltage $V_g$ measured at $T = 1.6$ K for device P3. A gapped region specified by a dark blue area is observed in the $V_g$ -$V_b$ plane where both band edges lie outside the bias window.

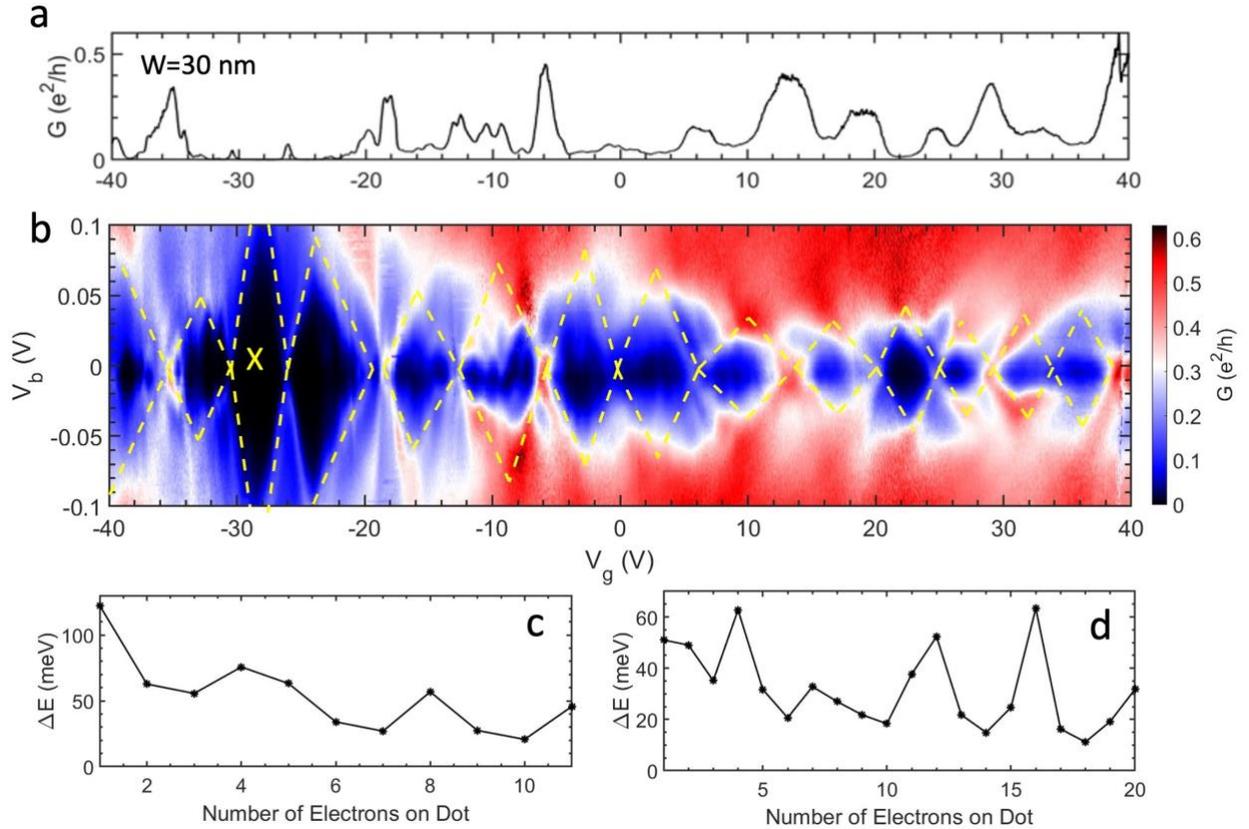

**FIG 3: Transport measurements of the narrowest constriction.** (**a**) Conductance vs back gate voltage, $V_g$, for device P4 with $W$=30 ± 5 nm measured at $T$=1.6 K. (**b**) Differential conductance as a function of bias, $V_b$, and back gate voltage, $V_g$, measured for device P4 exhibiting coulomb diamonds. The marks "X" represent the location of the charge neutrality point. The location of Coulomb peaks is matched with peak positions in Fig. 4. Addition energies of the QD, $\Delta E$, as a function of electron occupation number for the data displayed in (**c**) Fig. 3a, (**d**) Fig. S1a.

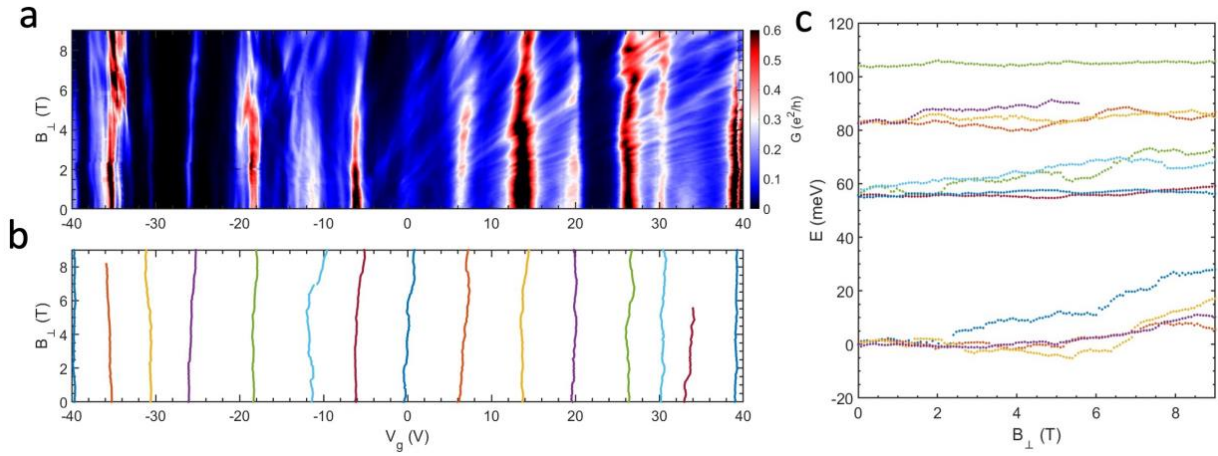

**FIG 4: Magnetic field measurements of bilayer graphene QD states.** (**a**) Conductance of device P4 as a function of back gate voltage, $V_g$, and magnetic field, $B$, showing the evolution of coulomb resonances with $B$. The map in (a) is taken at a fixed value of $V_b = 0$. (**b**) The locations of conductance peaks in $V_g$ and $B$ extracted from (a). (**c**) Single particle energy level dispersion of the QD as a function of perpendicular magnetic field $B\perp$ extracted from Fig. 4a for electrons. A four-level bunching of quantum energies is observed as expected. Overall, a slight dispersion is observed consistent with the small size of the QD.

*Supplementary Information*

**A: The size estimation of QD (Other methods):**

An estimate of the size of the QD can be also obtained using the extracted quantum energies. Using the band structure of bilayer graphene and assuming a square confinement potential, the diameter of the quantum dot can be estimated as $a = \sqrt{\frac{\hbar^2 \pi}{m^* \Delta E}}$ This calculation for the first two filled bands gives an estimate of the diameter to be 11.52 nm and 18.88 nm for the first and second cool downs, respectively.

Another method to determine the dot size is to treat the dot as a parallel plate capacitor. For a circular disc of diameter, $a$, $C_g = \frac{\varepsilon_0 \varepsilon_r \pi \frac{a^2}{4}}{D}$. Where $D$ is the distance between the plates (285 nm SiO$_2$ + 20 nm hBN) and $\varepsilon_r$=3.9 is the dielectric constant. Rewriting this equation, the diameter of the dot can be expressed as $a = \sqrt{\frac{4 C_g D}{\varepsilon_0 \varepsilon_r \pi}}$ yielding 12.97 nm and 18.46 nm for the first and second cool downs, respectively

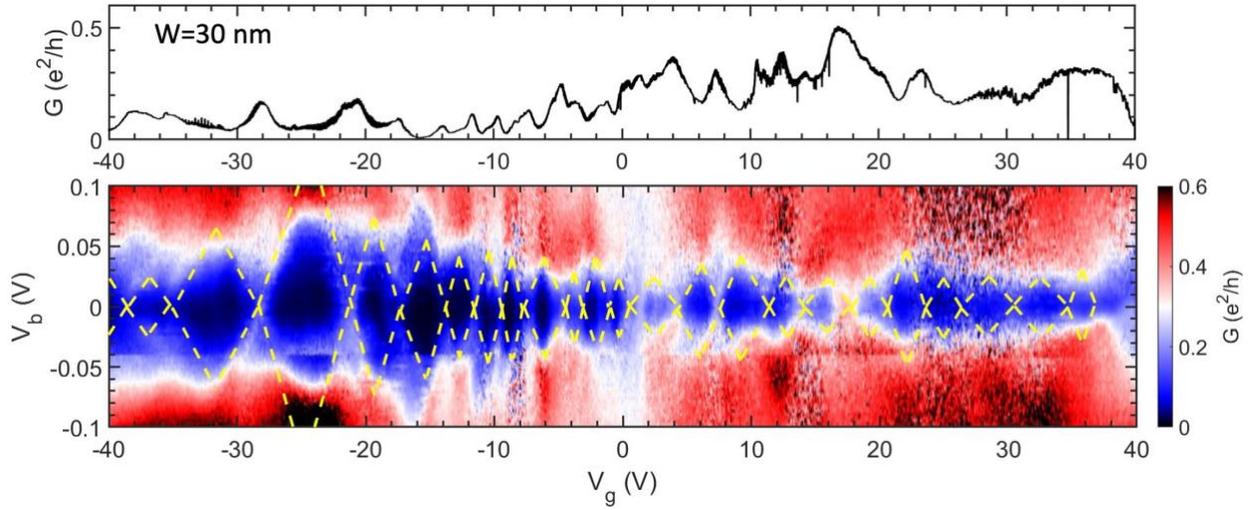

**FIG. S1: Transport measurements of the narrowest constriction after second cool down.** (**a**) Conductance vs back gate voltage, $V_g$, for device P4 with $W=30 \pm 5$ nm measured at $T=1.6$ K after the second cool down. (**b**) Differential conductance as a function of bias, $V_b$, and back gate voltage, $V_g$, measured for device P4 after second cool down. The marks "X" represent the location of charge neutrality point. The location of coulomb peaks is matched with peak positions in Fig. S2a.

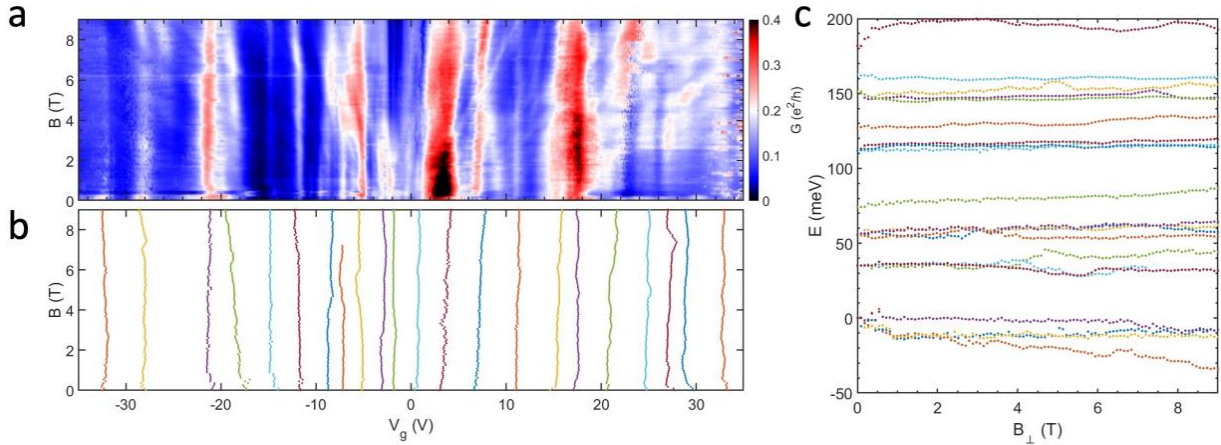

**FIG S2: Magnetic field measurements of device P3 in second cool down.** (**a**) Conductance of device P4 after second cool down as a function of back gate voltage, $V_g$, and magnetic field, $B$, showing the evolution of coulomb resonances with $B$. The map in (a) is taken at a fixed value of $V_b = 0$. (**b**) The locations of conductance peaks in $V_g$ and $B$ extracted from (a). (**c**) Single particle energy level dispersion of the QD as a function of perpendicular magnetic field $B\perp$ extracted from Fig. S2a for electrons. An approximated four level bunching of quantum energies is observed as expected. Overall, a slight dispersion is observed consistent with the small size of the QD.

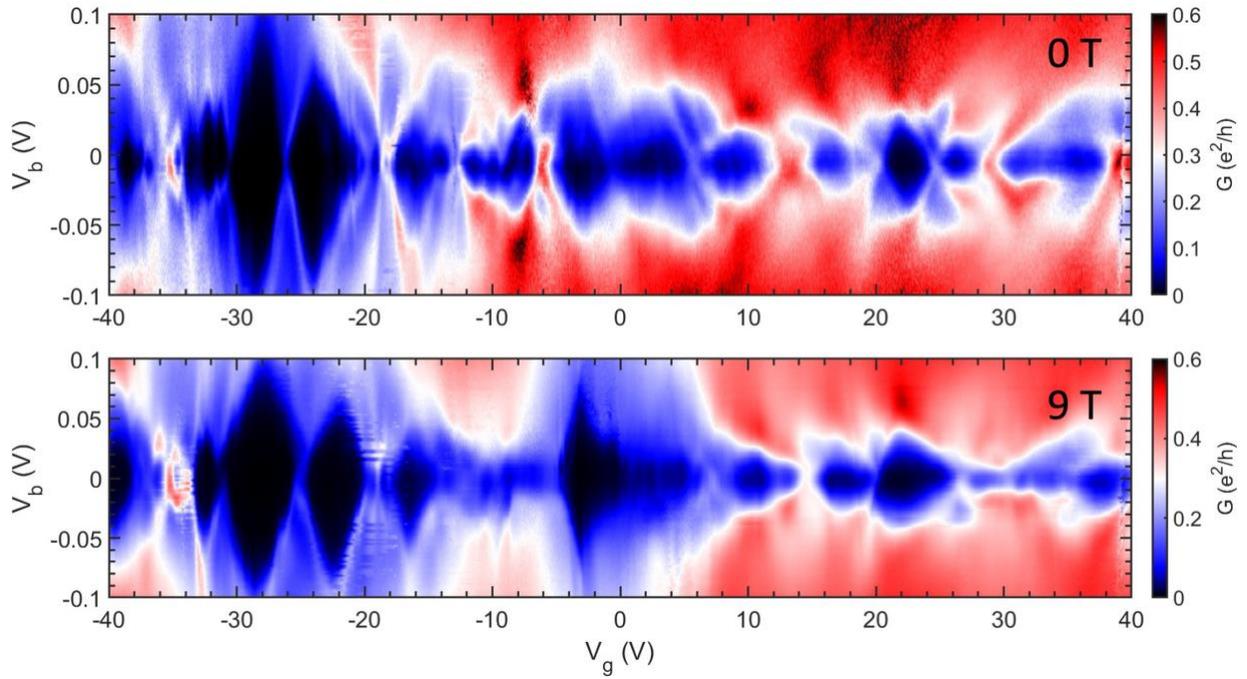

**FIG S3: Coulomb diamonds at 9T.** Comparison of Differential Conductance measured for device P4 (**a**) at zero field, and (**b**) at 9T.

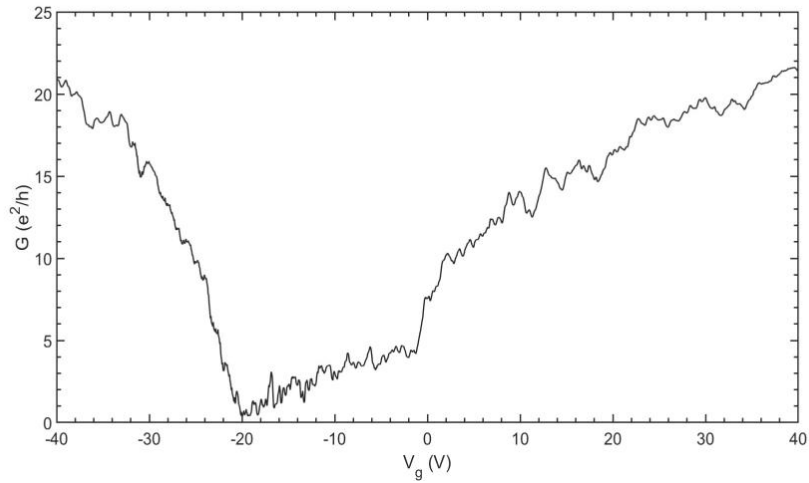

**FIG S4: Transport measurements of a graphene nano constriction.** Conductance of a graphene nano constriction (device P5) with a width of $W=70 \pm 5$ nm measured at $T= 1.6$ K. A region with suppressed conductance is observed around charge neutrality point indication of a transport gap as expected. The constriction is fabricated by first AFM etching of the flake followed by hBN encapsulation.

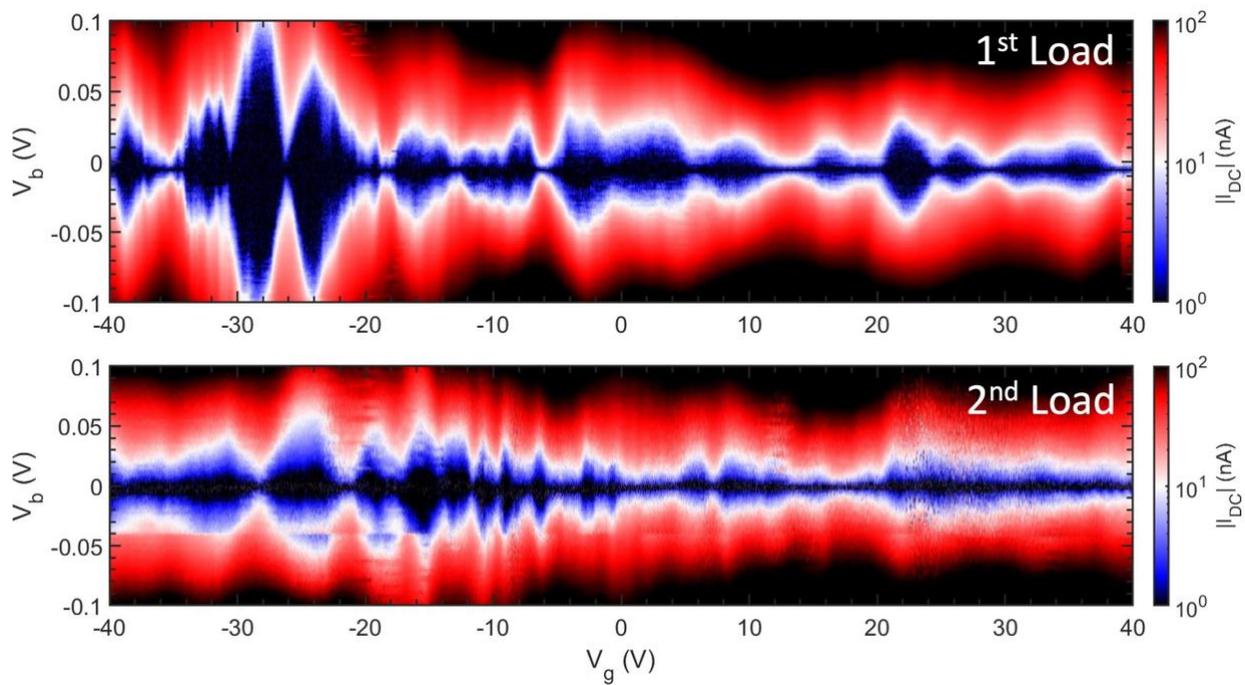

**FIG S5: DC technique.** Source-drain measurements of device P4 as a function of DC bias, $V_b$, and back gate voltage, $V_g$.